\documentclass[aps,prl,showpacs,amsmath,amsfonts,amssymb,twocolumn]{revtex4} 
\usepackage{amsthm}
\usepackage{bbm}
\usepackage{psfrag}
\usepackage[dvips]{graphicx}
\input epsf.sty

\newcommand\RR{{\mathbbm{R}}}
\newcommand\dd{{\mathrm{d}}}

\newcommand\ii{{\mathrm{i}}}

\DeclareMathOperator{\Dirac}{\delta}

\begin{document} 

\title{Microcanonical phase diagrams of short-range ferromagnets} 

\author{Michael Kastner} 
\email{kastner@sun.ac.za} 
\affiliation{National Institute for Theoretical Physics (NITheP), Stellenbosch 7600, South Africa} 

\author{Michel Pleimling} 
\email{michel.pleimling@vt.edu}
\affiliation{Department of Physics, Virginia Polytechnic Institute and State University, Blacksburg, Virginia 24061-0435, USA} 

\date{\today}
 
\begin{abstract}
A phase diagram is a graph in parameter space showing the phase boundaries of a many-particle system. Commonly, the control parameters are chosen to be those of the (generalized) canonical ensemble, such as temperature and magnetic field. However, depending on the physical situation of interest, the (generalized) microcanonical ensemble may be more appropriate, with the corresponding control parameters being energy and magnetization. We show that the phase diagram on this parameter space looks remarkably different from the canonical one. The general features of such a microcanonical phase diagram are investigated by studying two models of ferromagnets with short-range interactions. The physical consequences of the findings are discussed, including possible applications to nuclear fragmentation, adatoms on surfaces, and cold atoms in optical lattices.
\end{abstract}


\maketitle 

\begin{figure}[hb]
\psfrag{T}{{\small $T$}}
\psfrag{h}{{\small $h$}}
\psfrag{kritischer Punkt}{{\small critical point $(T_c,0)$}}
\vspace{-4mm}
\includegraphics[scale=0.44]{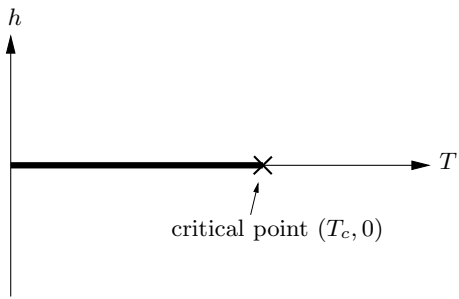}
\vspace{-1mm}
\caption{Phase diagram of a simple ferromagnet. The bold line marks the parameter values in the $(T,h)$ plane at which the Gibbs free energy density $g(T,h)$ is nonanalytic.}
\label{fig:PDcan}
\end{figure}

Cooperative effects can lead to remarkable properties of many-body systems, and the occurrence of a phase transition is a prime example of such an effect. At a phase transition, the macroscopic properties of a many-particle system change abruptly under variation of a control parameter. Typical examples of phase transitions are the evaporation of a liquid at temperatures above its boiling point, or the onset of a spontaneous magnetization in a ferromagnet below its Curie temperature. In a thermodynamic description, phase transitions are signaled by nonanalyticities of thermodynamic functions like the free energy density. For example, when discussing the phases and phase transitions of a ferromagnet, the Gibbs free energy density $g(T,h)$ as a function of the temperature $T$ and the external magnetic field $h$ is considered. The so-called phase diagram is obtained by plotting in parameter space the nonanalyticities of $g$, i.\,e., the points or lines in the $(T,h)$ plane at which this function is not infinitely many times differentiable. Whenever the parameters $T$ and $h$ are varied along a path crossing such a point or line of nonanalyticities, the system will undergo a phase transition. For a ferromagnet, such a phase diagram has a very simple appearance (see Fig.\ \ref{fig:PDcan}).

This is how a phase transitions appears on the macroscopic, thermodynamic level of description. On a more fundamental level, statistical physics provides a microscopic description underlying the thermodynamic one. Under suitable conditions on the interactions, thermodynamics can be recovered from the statistical description in the thermodynamic limit of infinite number of degrees of freedom. In equilibrium statistical physics, statistical weights are associated to the various microstates of a system, and the choice of these weights depends on the physical situation the system is in. Different physical situations are described by the so-called statistical ensembles. The microcanonical ensemble for example is appropriate for the description of an isolated system at fixed energy, whereas the canonical ensemble describes a system in equilibrium with an infinitely large heat bath of temperature $T$. For a suitable class of short-range interactions, both ensembles are known to give equivalent results in the thermodynamic limit (see \cite{Ruelle} for details). For long-range interactions, however, equivalence may be violated, in particular whenever a discontinuous phase transition takes place in the canonical ensemble \cite{TouElTur:04}.

But also in the case of short-range interactions when different ensembles yield equivalent results in the thermodynamic limit, an important difference remains, which we would like to discuss, taking again a simple ferromagnet as an example: Canonically, the temperature $T$ and the external magnetic field $h$ are the relevant control parameters. Microcanonically, however, this is no longer true. In a (generalized) microcanonical ensemble, the energy (density) $\varepsilon$ and the magnetization (density) $m$ are the natural control parameters corresponding to $T$ and $h$ \footnote{Relating the microcanonical and the canonical ensemble by means of a Legendre-Fenchel transform, $\varepsilon$ is thermodynamically conjugate to $\beta$, and $m$ to $-\beta h$, where $\beta=1/T$ is the inverse temperature and Boltzmann's constant has been set to unity.}. Working in this microcanonical ensemble, one would naturally ask: What does the phase diagram corresponding to the one in Fig.\ \ref{fig:PDcan} look like in the $(\varepsilon,m)$ plane? This diagram then could readily answer the question whether, upon variation of $\varepsilon$ and $m$ along a certain path, the system undergoes a phase transition or not. Remarkably, such a {\em microcanonical phase diagram}\/ looks very different from its canonical counterpart. Moreover, although the microcanonical ensemble is the most fundamental one among the statistical ensembles, little is known about these issues. To some extend this is also due to the fact that studies in the microcanonical ensemble, either analytically or numerically, are typically more demanding than in the canonical ensemble.  

In this Letter, we present two case studies which serve to illustrate the general properties of microcanonical phase diagrams of short-range interacting ferromagnets. The first example is the spherical model with nearest-neighbor interactions on a $d$-dimensional hypercubic lattice, for which analytical results are presented. Second, numerical results are reported for the Ising model with nearest-neighbor interactions on a two-dimensional square lattice. Remarkably, both models show {\em two}\/ distinct transition lines in the $(\varepsilon,m)$ plane which, for fixed magnetization $m$, may be crossed upon variation of the energy $\varepsilon$. In the conclusions, these unexpected results are discussed, in particular as what regards physical realizations of an ensemble where both $\varepsilon$ and $m$ are fixed.

{\em Spherical model.---}Introduced by Berlin and Kac \cite{BerKac:52} in 1952, this model was constructed to show a ferromagnetic phase transition while being exactly solvable. The degrees of freedom $\sigma_i\in\RR$ are associated to the sites of a $d$-dimensional hypercubic lattice. The energy of a microstate $\sigma=(\sigma_1,\dotsc,\sigma_N)$ of $N$ degrees of freedom is
\begin{equation}
	H(\sigma)=-J\sum_{\langle i,j\rangle}\sigma_i\sigma_j-h\sum_i \sigma_i,
\end{equation}
where $J>0$ is a coupling constant determining the strength of the exchange interaction. The angular brackets denote a summation over all pairs of nearest neighbors on the lattice. In addition, the $\sigma_i$ are required to satisfy the spherical constraint $\sum_i \sigma_i^2=N$, which accounts for the model's name. In the canonical ensemble, the spherical model is exactly solvable in the thermodynamic limit for arbitrary spatial dimension $d$, and a transition from a ferromagnetic phase at low temperatures to a paramagnetic phase at high temperatures occurs for $d\geqslant3$.

\begin{figure}[hb]\center
\psfrag{e}{\small $\varepsilon$}
\psfrag{m}{\small $m$}
\psfrag{1}{\scriptsize $1$}
\psfrag{2}{\scriptsize $2$}
\psfrag{3}{\scriptsize $3$}
\psfrag{0.5}{\scriptsize $0.5$}
\psfrag{1.0}{\scriptsize $\!1.0$}
\psfrag{-}{\scriptsize $\!-$}
\psfrag{I}{}
\psfrag{II}{}
\psfrag{III}{}
\vspace{-4mm}
\includegraphics[scale=1.0]{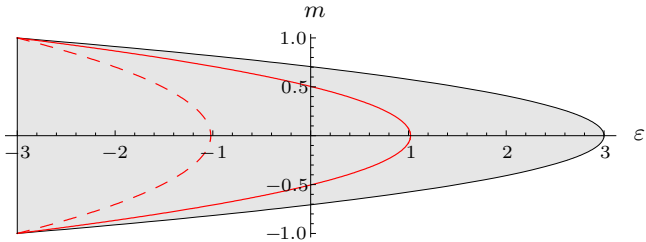}
\vspace{-1mm}
\caption{\label{fig:phasediagram}
(Color online) Microcanonical phase diagram of the spherical model on a three-dimensional cubic lattice. The microcanonical entropy $s$ is defined only within the gray shaded region in the $(\varepsilon,m)$ plane. Within each of the three gray shaded subregions, the entropy is analytic, but not so on their boundaries which are given by Eqs.\ \eqref{eq:epm} and \eqref{eq:ad}.
}
\end{figure}

Starting point for a calculation in the microcanonical ensemble is the density of states as a function of energy $\varepsilon$ and magnetization $m$,
\begin{multline}
\Omega_N(\varepsilon,m)=\int_{\RR^N}\dd\sigma\,\Dirac[N\varepsilon-H(\sigma)]\\
\times\Dirac[Nm-M(\sigma)]\,\Dirac\bigg(N-\sum_{i=1}^N\sigma_i^2\bigg),
\end{multline}
where $M(\sigma)=\sum_{i=1}^N \sigma_i$ yields the total magnetization of a microstate. An analytic calculation of $\Omega_N$ is reported in \cite{Behringer:05}, but the saddle point analysis proposed in that paper works only within a certain range of $\varepsilon$ and $m$ values. This can be seen by starting from Eqs.\ (25) and (26) of Ref.\ \cite{Behringer:05} and, by explicitely performing two of the integrations, rewriting the density of states for large $N$ as
\begin{widetext}
\begin{equation}
\Omega_N(\varepsilon,m)\sim \frac{N^{(N-6)/2}}{\Gamma[(N-3)/2]}\int_{a-\ii\infty}^{a+\ii\infty}\frac{\dd z}{2\pi}
\sqrt{\frac{1-zdJ}{\left[1-m^2+z(\varepsilon+m^2dJ)\right]^5}} \exp\!\bigg[-\frac{N}{2\pi^d}\int_{[0,\pi)^d}\dd^d\varphi\,\ln\!\bigg(\frac{1-zJ\sum_{j=1}^d\cos\varphi_j}{1-m^2+z(\varepsilon+m^2dJ)}\bigg)\!\bigg].\label{eq:Omega_1int}
\end{equation}
\end{widetext}
A derivation of this result will be given elsewhere. Due to the logarithm in \eqref{eq:Omega_1int}, the integrand of the $z$-integration has two branch cuts on the real line. Apart from these branch cuts, the integrand is holomorphic, and the contour of integration can be deformed freely, as long as it does not cross the cuts. For an asymptotic evaluation of the integral in the large-$N$-limit by means of the method of steepest descent, the path of integration in the complex plane is deformed such that its imaginary part becomes zero. In this limit, the value of the integral is given by the integrand of the $z$-integration in \eqref{eq:Omega_1int} evaluated at the maximum along that path. Depending on the values of $\varepsilon$ and $m$, this maximum may either be a saddle point of the exponent in \eqref{eq:Omega_1int} (and in this case the analysis in \cite{Behringer:05} is valid), or located at one of the end points of the branch cuts. The transition between these two types of behavior accounts for nonanalyticities of the microcanonical entropy in the thermodynamic limit \footnote{Note that the free energy $g(\beta,h)$ is the Legendre-Fenchel transform of $s(\varepsilon,m)$.}, 
\begin{equation}\label{eq:sem}
s(\varepsilon,m)=\lim_{N\to\infty}\frac{1}{N}\ln\Omega_N(\varepsilon,m),
\end{equation}
and hence for the occurrence of phase transitions \footnote{This mechanism of how nonanalyticities emerge from an asymptotic evaluation of an integral is very similar to the one observed by Farago [J.\ Stat.\ Phys.\ {\bfseries 107}, 781 (2002)] for nonequilibrium processes.}. Note that, contrary to what has been conjectured in \cite{KaSchneSchrei:07,KaSchneSchrei:08}, the entropy $s$ of the spherical model is found to be a concave function on its entire domain. By means of an asymptotic analysis of Eqs.\ \eqref{eq:Omega_1int} and \eqref{eq:sem}, the values of $\varepsilon$ and $m$ for which $s(\varepsilon,m)$ becomes nonanalytic can be computed, yielding
\begin{equation}\label{eq:epm}
\varepsilon_\pm(m)=-dJ\frac{m^2+a_d\left[m^2\pm(1-m^2)\right]}{1+a_d}
\end{equation}
with
\begin{equation}\label{eq:ad}
a_d=\int_{[0,\pi)^d}\frac{\dd^d\varphi}{\pi^d}\,\frac{\sum_{j=1}^d\cos\varphi_j}{d-\sum_{j=1}^d\cos\varphi_j}.
\end{equation}
Plotting the two curves $\varepsilon_\pm(m)$ in the $(\varepsilon,m)$ plane, the microcanonical phase diagram of the spherical model is obtained (see Fig.\ \ref{fig:phasediagram} for a plot of the $d=3$ case).

This microcanonical phase diagram looks remarkably different from its canonical counterpart in Fig.\ \ref{fig:PDcan}, and even an experienced statistical physicist, we suspect, would have had problems predicting its shape. Varying, for example, the energy $\varepsilon$ while keeping fixed the magnetization at any value of $m$, one typically crosses {\em two}\/ transition lines in the phase diagram, therefore observing two phase transitions, signaled by kinks in the specific heat. Similarly, four transition lines are crossed upon variation of the magnetization while keeping the energy $\varepsilon$ fixed at any value $-d<\varepsilon<\varepsilon_+(0)$, while two transition lines are crossed for energies $\varepsilon_+(0)<\varepsilon<\varepsilon_-(0)$.

The dashed line in Fig.\ \ref{fig:phasediagram} corresponds to a transition from a ferromagnetic to a paramagnetic phase, and the region to the left of this line is the coexistence region. When switching to the canonical ensemble by means of a Legendre-Fenchel transform, the entire coexistence region is mapped onto the transition line in the canonical phase diagram. The solid line in Fig.\ \ref{fig:phasediagram}, in contrast, has no counterpart in Fig.\ \ref{fig:PDcan}, mainly due to the fact that this transition occurs at negative microcanonical inverse temperatures $ds/d\varepsilon$. So what kind of phase is then found to the right of this second transition line? At least for vanishing magnetization $m$ one can argue that, upon crossing this line, a transition to an antiferromagnetic phase takes place \footnote{This follows from a mapping of the model with negative coupling J at positive temperatures onto the same model with positive coupling at negative temperatures.}. For $m\neq0$, however, an interpretation of the transition is more difficult since $m$ is not an order parameter of the antiferromagnetic transition.

{\em Two-dimensional Ising model.---}The Ising model is arguably the most studied model in the theory of phase transitions, serving as a test case also for our aim of computing the microcanonical phase diagram. Its energy function is formally equivalent to that of the spherical model on a two-dimensional square lattice,
\begin{equation}
	H(\sigma_1,\dots,\sigma_N)=-J\sum_{\langle i,j\rangle}\sigma_i\sigma_j-h\sum_i \sigma_i,
\end{equation}
but the degrees of freedom $\sigma_i\in\{-1,+1\}$ take on only discrete values. Again, the angular brackets denote a summation over all pairs of nearest neighbors on the lattice. For vanishing external field $h$ and in the thermodynamic limit, this model is known to undergo a phase transition from a ferromagnetic phase at low temperatures to a paramagnetic phase at high temperatures, taking place at a critical inverse temperature $\beta_\text{c}=\ln(1+\sqrt{2})/2$. The analytic solution for the free energy density was obtained by Onsager in 1944 \cite{Onsager:44}. For $h\neq0$, however, no analytic solution is known. In the microcanonical framework, this corresponds to the fact that analytic results exist only for some regions in the $(\varepsilon,m)$ plane, but not for all. Consequently, we will resort to numerical methods in order to compute the microcanonical phase diagram.

We use a Monte Carlo histogram method discussed in \cite{PleimHu:01} in the context of the Ising model with fixed magnetization. For square Ising systems composed of $N = L^2$ lattice sites, we compute the density of states $\Omega_N(\varepsilon,m_0)$ for a clamped value of the magnetization $m = m_0$. In a microcanonical analysis, quantities of interest are directly derived from $\Omega_N(\varepsilon,m_0)$ or, equivalently, from the microcanonical entropy $s_N(\varepsilon,m_0) = \ln \Omega_N(\varepsilon,m_0)/N$. The microcanonical specific heat at fixed magnetization $m = m_0$, for instance, is given by $c(\varepsilon) = - [ds/d\varepsilon ]^2/[d^2s/d\varepsilon^2]$. It is this quantity that we use for our investigation of the microcanonical phase diagram of the two-dimensional Ising model, mainly by looking for peaks in the specific heat which can be viewed as finite-system precursors of nonanalyticities occurring in the thermodynamic limit.

\begin{figure}[ht]
\centerline{\epsfxsize=3.25in\ \epsfbox{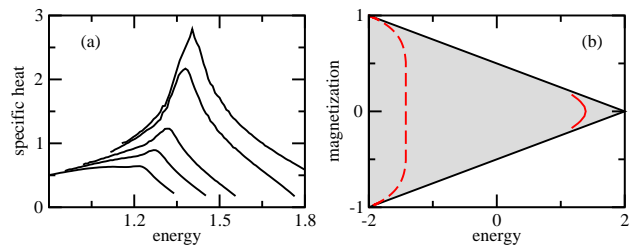}}
\vspace{-2mm}
\caption{(Color online) (a) Microcanonical specific heat as a function of energy $\varepsilon$ for fixed values of the magnetization $m=m_0$. From top to bottom: $m_0 = 0, 0.05, 0.1, 0.125, 0.15$. The peaks reveal the presence of a phase transition line at relatively high energies. The data shown have been obtained for systems composed of $100 \times 100$ spins. (b) Resulting microcanonical phase diagram of the two-dimensional Ising model. The dashed line is the line of spontaneous magnetization separating the ferromagnetic phase at low energies from the paramagnetic phase. The solid line is new and should be compared with the corresponding line in the microcanonical phase diagram of the spherical model in Fig.\ \ref{fig:phasediagram}.
\label{fig3}
}
\vspace{-4mm}
\end{figure}

Fig.\ \ref{fig3} summarizes our results for systems composed of $100 \times 100$ spins.%
When looking at the microcanonical specific heat at a certain (for practical reasons not too large) value of the fixed magnetization, we observe {\em two}\/ peaks. One peak occurs at lower energies, signaling the crossing of the coexistence line which separates the ferromagnetic phase from the paramagnetic phase (not shown) \cite{PleimHu:01,HuPleim:02,Kastner:02}. A second peak is observed for much larger energies, as shown in Fig.\ \ref{fig3}(a). The positions of these peaks shift to smaller energies when the fixed value of the magnetization increases. We have plotted the peak positions in the $(\varepsilon, m)$ plane in Fig.\ \ref{fig3}(b), yielding the microcanonical phase diagram of the Ising model. The phase diagram of the Ising model strikingly resembles the corresponding phase diagram of the spherical model shown in Fig. \ref{fig:phasediagram}. In both cases there is a range of magnetizations for which lines at fixed energy cross two different transition lines. Note that for the Ising model, we are not able to determine whether the new line extends all the way down to the ground state, as this line rapidly closes in on the boundary of the microcanonical entropy's support.

In order to check the robustness of our findings for the Ising model, we also studied larger systems with up to $600 \times 600$ spins. We observe that the peaks increase in height and get sharper for increasing system sizes, as expected for a phase transition. The positions of the peaks shift slightly towards larger values of $\varepsilon$ when increasing the system size, but this shift is so small that it remains within the thickness of the line on the scale of Fig.\ \ref{fig3}(b). For very large systems, it is extremely difficult to obtain the high quality data needed for a microcanonical analysis, and we were not able to make a quantitative study of the change in peak height and position.

{\em Discussion.---}The phase diagrams discussed above represent a physical situation in which both energy $\varepsilon$ and magnetization $m$ can be controlled externally. Considering the spherical model or the Ising model properly as models of ferromagnets, control of the energy may well be imagined in an experimental set-up energetically isolated from the environment. Direct control of the magnetization, however, appears difficult---if not impossible---to achieve. But ferromagnetic spin models have a wide range of applications, going well beyond the modeling of ferromagnetic materials, both in classical and quantum physics. Following Lee and Yang, the Ising model can be mapped onto a lattice gas, in which the magnetization within the first model formally corresponds to the particle density within the latter \cite{LeeYangII:52}. Control of the particle density is of course an experimentally realistic scenario, and in this situation the microcanonical phase diagrams of ferromagnetic models can provide relevant information. Examples include the Ising model as a model of nuclear matter fragmentation \cite{Carmona_etal:98} or as a model of adatoms on a crystal surface \cite{MuelSel:99}. Cold atoms in an optical lattice are another possible experimental realization: After switching off the cooling, total energy and number of atoms are conserved to a very good degree, rendering appropriate a description as a lattice gas in the microcanonical ensemble. The interactions between the atoms can be tuned via Feshbach resonances, allowing to realize, among others, Ising-type interactions \cite{Duan_etal:03}.

A comment is in order on the short-range nature of the interactions in the two examples discussed. Although we believe that the qualitative behavior of the microcanonical phase diagram should not be restricted to nearest-neighbor interactions, it surely does not extend to ferromagnets with long-range interactions. This becomes obvious when considering for example the mean-field $\varphi^4$ model. Although this model undergoes a ferromagnetic transition, a calculation of the microcanonical entropy $s(\varepsilon,m)$ as a function of energy and magnetization yields a smooth function \cite{HaKa:05,HaKa:06}. Therefore, the microcanonical phase diagram of the mean-field $\varphi^4$ model in the $(\varepsilon,m)$ plane does not show any transition lines at all. This can be seen as a consequence of the nonequivalence of the microcanonical and the canonical ensemble in this long-range interacting model. Apart from short-range interactions, we also expect an upper bound on the energy per particle to be essential for the observed behavior. This is usually the case for spin models, often allowing for negative microcanonical temperatures to occur.

{\em Conclusions.---}We have computed microcanonical phase diagrams in the parameter space of energy and magnetization for two ferromagnetic models. The diagrams look remarkably different from the corresponding canonical ones in the $(T,h)$ plane, with the consequence that, when controlling $\varepsilon$ and $m$ in a microcanonical setting, the physical behavior differs significantly from the canonical situation in which $T$ and $h$ are controlled. For both models investigated, the microcanonical phase diagrams are qualitatively similar to each other, and we expect this to extend to short-range ferromagnetic spin models more generally. Finally, we have pointed out physical applications of microcanonical phase diagrams within the lattice gas interpretation of the Ising model, including nuclear fragmentation, adatoms on surfaces, and cold atoms in optical lattices.


\bibliography{MicroPhaseDiagram.bib}

\end{document}